\documentstyle[aps, epsf, psfig]{revtex}

\newcommand{\na}{\nabla}

\newcommand{\ep}{\epsilon}

\newcommand{\vphi}{\varphi}

\newcommand{\lraw}{\longrightarrow}

\newcommand{\pa}{\partial}

\newcommand{\td}{\tilde}
\newcommand{\sla}[1]{\slash\!\!\! #1}
\begin{document}

\draft
\begin{center}
{\Large\bf Isospin Breaking and $\omega\rightarrow\pi^+\pi^-$
Decay}\\[5mm]
Xiao-Jun Wang\footnote{E-mail address: wangxj@mail.ustc.edu.cn} \\
{\small
Center for Fundamental Physics,
University of Science and Technology of China\\
Hefei, Anhui 230026, P.R. China} \\
Mu-Lin Yan\\
{\small CCST(World Lad), P.O. Box 8730, Beijing, 100080, P.R. China}\\
{\small and}\\
{\small  Center for Fundamental Physics,
University of Science and Technology of China\\
Hefei, Anhui 230026, P.R. China}\footnote{mail
address}
\end{center}
\begin{abstract}
We study $\omega\rightarrow\pi^+\pi^-$ decay up to including all
orders of the chiral expansion and one-loop level of mesons in formlism of
chiral constituent quark model. This G-parity forbidden decay is caused by
$m_u\neq m_d$ and electromagnetic interaction of mesons. We illustrate
that in the formlism both nonresonant contact interaction and $\rho$
resonance exchange contribute to this process, and the contribution from
$\rho$ resonance exchange is dominant. We obtain that transition matrix
element is $<~\rho|H_{\rho\omega}|\omega>=[-(3956\pm 280)-(1697\pm
130)i]$MeV$^2$, and isospin breaking parameter is $m_d-m_u=3.9\pm 0.22$MeV
at energy scale $\mu\sim m_\omega$.
\end{abstract}

\section{Introduction}

The light current quark masses are baisc input of quantum 
chromodynamics(QCD). The inequality of the light quark masses, especially,
$m_u\neq m_d$, breaks the isospin symmetry or charge
symmetry\cite{Miller90}. This breaking of isospin symmetry induces various
measurable physics processes such as $\pi^0-\eta$, $\Lambda-\Sigma^0$
mixing and $\omega\rightarrow\pi^+\pi^-$ decay etc. In this paper, we will
focus on the $\omega\rightarrow\pi^+\pi^-$ decay, which is considered as
the important source of charge symmetry breaking in nuclear physics.

In ref.\cite{Wang00}(we will quote it as ${\bf I}$ hereafter) we have
shown that the chiral expansion at vector meson energy region converges
slowly. Therefore, a well-defined effective field theory describing the
physics in this energy region must be available for calculation on high
order terms of the chiral expansion and meson loops. Obviously, method of
chiral perturbative theory(ChPT)\cite{GL85a} is impractical to cupature
the high order term contribution because the number of free parameter
increases very rapidly as perturbative order raising. In 
${\bf I}$, following the spirit of Manohar-Georgi(MG) model\cite{GM84}, we
have constructed chiral constituent quark model(ChCQM) including lowest
vector meson resonances. The advatanges of this approach are that high
order contribution of the chiral expansion and $N_c^{-1}$ expansion can be
calculated consistently, and only fewer free parameters are required.
Low energy limit and unitarity of the model are also examined
successfully. In particular, it is a attractive property that although the
leading order theoretical prediction does not match with experimental
data, larger contribution from high order of the chiral expansion and
pseudoscalar meson one-loop corrects theoretical prediction close to data
very much. It is just the characteristic of the chiral expansion in this
energy region. Therefore, in this the present paper, we also need to
calculate $\omega\rightarrow\pi^+\pi^-$ decay to include all high order
contribution and pseudoscalar meson one-loop correction. 

This research is also motivated by the following reasons:

i) In the most recent references\cite{CB87,Connell97,Bernicha94}, the
$\omega\rightarrow\pi^+\pi^-$ decay was treated as being dominant via
$\rho$-resonance exchange, and the direct $\omega\pi^+\pi^-$ coupling
is neglected. It has been pointed out in ref.\cite{Maltman96} that the
neglect of $\omega$ ``direct'' coupling to $\pi^+\pi^-$ is not valid. It
can be naturally understood since $\pi\pi$ can make up of vector-isovector
system, whose quantum numbers are same to $\rho$ meson. Thus in an
effective lagrangian based on chiral symmetry, every $\rho$ field can be
replaced by $\pi\pi$ and does not conflict with symmetry. Although authors
of ref.\cite{Maltman96} also pointed out that the present experimental
data still can not be used to separate ``direct'' $\omega\pi\pi$ coupling
from $\omega-\rho$ mixing contribution in model-independent way, it is
very interesting to perform a theorectical investigation on ``direct''
$\omega\pi\pi$ coupling contribution. We will show that the contribution
from interfernce of ``direct'' $\omega\pi\pi$ coupling and $\omega-\rho$
is about 15$\%$. Thus ``direct'' $\omega\pi\pi$ coupling can not be
neglected indeed.  

ii) The present study involves the investigation of $\rho^0-\omega$
mixing, which has been an active subject[5--14]. The mixing amplitude for
on-mass-shell vector mesons has been observed directly in the measurement
of the pion form-factor in the time-like region from the process
$e^+e^-\rightarrow\pi^+\pi^-$\cite{OC98,Barkov85}. For roughly twenty
years, $\rho^0-\omega$ mixing amplitude was assumed constant or momentum
independt even if $\rho$ and $\omega$ have the space-like momenta, far
from the on-shell point. Several years ago, this assumption was firstly
questioned by Goldman {\sl et. al.}\cite{GHT}, and the mixing amplitude
was found to be significantly momentum dependent within a simple quark
loop model. Subsequently, various authors have argued such momentum
dependence of the $\rho^0-\omega$ mixing amplitude by using various
approaches\cite{Momenta,OC94}. In particular, the authors of
ref.\cite{OC94} has pointed out that $\rho^0-\omega$ mixing amplitude must
vanish at $q^2=0$(where $q^2$ denotes the four-momentum square of the
vector mesons) within a broad class of model. This point will be also
exmined in ChCQM.

iii) It has been known that $\omega\rightarrow\pi^+\pi^-$ decay amplitude
receive the contribution from two sources: isospin symmetry breaking due
to $u-d$ quark mass difference and electromagnetic interaction. In {\bf
I} we have shown that VMD\cite{Sakurai69} in meson physics is natural
consequence of the present formlism instead of input. The
vector~$\rightarrow e^+e^-$ decays are also predicted successfully.
Therefore, the dynamics of electromagnetic interactions of mesons has been
well established, and the calculation for $\omega\rightarrow\pi^+\pi^-$
decay from the transition
$\omega\rightarrow\gamma\rightarrow\rho\rightarrow\pi\pi$
and ``direct'' $\omega\rightarrow\gamma\rightarrow\pi\pi$ is
straightforward. In this the present paper, we will pay our attention to
isospin breaking due to $m_u\neq m_d$. It is another purpose of this paper
to determine isospin breaking parameter $\delta m_q\equiv m_d-m_u$ via
$\omega\rightarrow\pi^+\pi^-$ decay. This parameter is urgently wanted by
determination of light quark mass ratios. 

The contents of the paper are
organized as follows. In sect. 2 we  review the basic notations of the
chiral constituent quark model with the lowest vector meson resonances. In
sect. 3, the tree level effective lagrangian, which including all order
contribution of the chiral expansion, is obtained. The pseudoscalar meson
one-loop corrections are calculated in sect. 5. In sect. 6, the formulas
and numerical results of $\omega-\rho^0$ mixing amplitude and
$\omega\rightarrow\pi^+\pi^-$ are given. The sect. 7 is devoted to a brief
summary.

\section{Chiral Constituent Quark Model with Vector Meson}

The simplest version of chiral quark model which was originated by 
Weinberg\cite{Wein79}, and developed by Manohar and Georgi\cite{GM84}
provides a QCD-inspired description on the simple constituent quark model.
In view of this model, in the energy region between the chiral symmetry
spontaneously broken (CSSB) scale and the
confinement scale ($\Lambda_{QCD}\sim 0.2-0.3 GeV$), the dynamical field
degrees of freedom are constituent quarks(quasi-particle of quarks),
gluons and Goldstone bosons associated with CSSB(these Goldstone bosons
correspond to lowest pseudoscalar octet). In this quasiparticle
description, the effective coupling between gluon and quarks is small and
the important interaction is the coupling between quarks and Goldstone
bosons. In {\bf I} we have further included the lowest
vector meson resonances into this formlism. At chiral limit, this model is
parameterized by the following chiral constituent quark lagrangian
\begin{eqnarray}\label{2.1}
{\cal L}_{\chi}&=&i\bar{q}(\sla{\pa}+\sla{\Gamma}+
  g_{_A}{\slash\!\!\!\!\Delta}\gamma_5-i\sla{V})q-m\bar{q}q 
   +\frac{F^2}{16}<\nabla_\mu U\nabla^\mu U^{\dag}>
   +\frac{1}{4}m_0^2<V_\mu V^{\mu}>.
\end{eqnarray}
Here $<...>$ denotes trace in SU(3) flavour space,
$\bar{q}=(\bar{q}_u,\bar{q}_d,\bar{q}_s)$ are constituent quark fields.
$V_\mu$ denotes vector meson octet and singlet, or more convenience, due
to OZI rule, they are combined into a singlet ``nonet'' matrix
\begin{equation}\label{2.2}
   V_\mu(x)={\bf \lambda \cdot V}_\mu =\sqrt{2}
\left(\begin{array}{ccc}
       \frac{\rho^0_\mu}{\sqrt{2}}+\frac{\omega_\mu}{\sqrt{2}}
            &\rho^+_\mu &K^{*+}_\mu   \\
    \rho^-_\mu&-\frac{\rho^0_\mu}{\sqrt{2}}+\frac{\omega_\mu}{\sqrt{2}}   
            &K^{*0}_\mu   \\
       K^{*-}_\mu&\bar{K}^{*0}_\mu&\phi_\mu
       \end{array} \right).
\end{equation}
The $\Delta_\mu$ and $\Gamma_\mu$ are defined as follows,
\begin{eqnarray}\label{2.3}
\Delta_\mu&=&\frac{1}{2}\{\xi^{\dag}(\pa_\mu-ir_\mu)\xi
          -\xi(\pa_\mu-il_\mu)\xi^{\dag}\}, \nonumber \\
\Gamma_\mu&=&\frac{1}{2}\{\xi^{\dag}(\pa_\mu-ir_\mu)\xi
          +\xi(\pa_\mu-il_\mu)\xi^{\dag}\},
\end{eqnarray}
and covariant derivative are defined as follows
\begin{eqnarray}\label{2.4}
\nabla_\mu U&=&\pa_\mu U-ir_\mu U+iUl_\mu=2\xi\Delta_\mu\xi,
  \nonumber \\
\nabla_\mu U^{\dag}&=&\pa_\mu U^{\dag}-il_\mu U^{\dag}+iU^{\dag}r_\mu
  =-2\xi^{\dag}\Delta\xi^{\dag},
\end{eqnarray}
where $l_\mu=v_\mu+a_\mu$ and $r_\mu=v_\mu-a_\mu$ are linear combinations
of external vector field $v_\mu$ and axial-vector field $a_\mu$, $\xi$
associates with non-linear realization of spontanoeusly broken global
chiral symmetry introduced by Weinberg\cite{Wein68}. This realization is
obtained by specifying the action of global chiral group $G=SU(3)_L\times
SU(3)_R$ on element $\xi(\Phi)$ of the coset space $G/SU(3)_{_V}$:
\begin{equation}\label{2.5}
\xi(\Phi)\rightarrow
g_R\xi(\Phi)h^{\dag}(\Phi)=h(\Phi)\xi(\Phi)g_L^{\dag},\hspace{0.5in}
 g_L, g_R\in G,\;\;h(\Phi)\in H=SU(3)_{_V}.
\end{equation}
Explicit form of $\xi(\Phi)$ is usual taken
\begin{equation}\label{2.6}   
\xi(\Phi)=\exp{\{i\lambda^a \Phi^a(x)/2\}},\hspace{1in}
U(\Phi)=\xi^2(\Phi),
\end{equation}
where the Goldstone boson $\Phi^a$ are treated as pseudoscalar meson
octet. The compensating $SU(3)_{_V}$ transformation $h(\Phi)$
defined by eq.(~\ref{2.3}) is th wanted ingredent for a non-linear
realization of G. In practice, we shall be interested in transformations
of $V_\mu,\;\Delta_\mu,\;\Gamma_\mu$ and constituent quark fields under
$SU(3)_{_V}$. The $q,\;\bar{q}$ transform as matter fields of
SU(3)$_{_V}$, 
\begin{equation}\label{2.7}
  q\lraw h(\Phi)q, \hspace{1in} \bar{q}\lraw \bar{q}h^{\dag}(\Phi).
\end{equation}
The vector meson fields transform homogeneously under SU(3)$_{_V}$
\begin{equation}\label{2.8}
 V_\mu\lraw h(\Phi)V_\mu h^{\dag}(\Phi),
\end{equation}
which was suggested by Weinberg\cite{Wein68} and developed further by
Callan, Coleman et. al.\cite{CCWZ69}. Since under local G, the expilcit
transformations of external vector and axial-vector fields are
\begin{eqnarray}\label{2.9}
&&l_\mu\equiv v_\mu-a_\mu \rightarrow g_L(x)l_\mu g_L^{\dagger}(x)
           +ig_L(x)\partial_\mu g_L^{\dagger}(x), \nonumber \\
      &&r_\mu\equiv v_\mu+a_\mu \rightarrow g_R(x)r_\mu g_R^{\dagger}(x)
           +ig_R(x)\partial_\mu g_R^{\dagger}(x),
\end{eqnarray}
$\Delta_\mu$ is SU(3)$_{_V}$ is invariant field gradients and $\Gamma_\mu$
transforms as field connection of SU(3)$_{_V}$
\begin{equation}\label{2.10}
\Delta_\mu\lraw h(\Phi)\Delta_\mu h^{\dag}(\Phi), \hspace{0.8in}
\Gamma_\mu\lraw h(\Phi)\Gamma_\mu h^{\dag}(\Phi)+h(\Phi)\pa_\mu
  h^{\dag}(\Phi).
\end{equation}
Thus the lagrangian(~\ref{2.1}) is invariant under $G_{\rm global}\times
G_{\rm local}$.

The several remarks are need here. 1) Note that there is no kinetic term
for vector meson fields in ${\cal L}_\chi$. Therefore, in this formlism
the vector mesons are treated as composited fields of constituent equarks
instead of fundamental fields. The dynamics of vector meson resonances
will be generated via loop effects of constituent quarks. 2) Note that
there is kinetic term of pseudoscalar mesons in ${\cal L}_\chi$. This
is different from some other chiral quark models, in which
there is no such term.  Existing of this kinetic term is consistent with
basic assumption of our model, because in this energy region, the
dynamical field degrees of freedom are both constituent quarks and
Goldstone bosons associated with
CSSB. 3) In ${\cal L}_\chi$ the parameter $g_A\simeq 0.75$ is determined
by $\beta$ decay of neutron. It has been pointed out in {\bf I} that this
value has included effects of intermediate axial-vector resonances
exchanges at low energy. In addition, the constituent quark mass parameter
$m\simeq 480$MeV has been fitted in {\bf I} via low energy limit of the
model. Such large value is requireed by convergence of chiral expansion at
vector meson energy scale. 

In this paper, we must go beyond chiral limit for obtaining isospin
breaking results. The light current quark matrix ${\cal M}={\rm
diag}\{m_u,m_d,m_s\}$ can be usually included into external scalar fields,
i.e., $\td{\chi}=s+ip$, where $s=s_{ext}+{\cal M}$, $s_{\rm ext}$ and $p$
are scalar and pseudoscalar external fields respectively. The chiral
transformation for $\td{\chi}$ is
\begin{equation}\label{2.11}
\td{\chi}\lraw g_R\td{\chi}g_L^{\dag}.
\end{equation} 
Thus together with $\xi$ and $\xi^{\dag}$, $\td{\chi}$ and
$\td{\chi}^{\dag}$ can form SU(3)$_{_V}$ invariant scalar source
$\xi^{\dag}\td{\chi}\xi^{\dag}+\xi\td{\chi}^{\dag}\xi$ and pseudoscalar
source $(\xi^{\dag}\td{\chi}\xi^{\dag}-\xi\td{\chi}^{\dag}\xi)\gamma_5$.
Then current quark mass dependent lagrangian is written
\begin{equation}\label{2.12}
-\frac{1}{2}\bar{q}(\xi^{\dag}\td{\chi}\xi^{\dag}
+\xi\td{\chi}^{\dag}\xi)q-\frac{\kappa}{2}
 \bar{q}(\xi^{\dag}\td{\chi}\xi^{\dag}-\xi\td{\chi}^{\dag}\xi)
 \gamma_5q.
\end{equation}
The above lagrangian will return to QCD lagrangian $\bar{\psi}{\cal
M}\psi$ in absence of pseudoscalar mesons at high energy. So that there is
a free parameter $\kappa$ which can not be determined by symmetry alonely.
From viewpoint of phenomenology, this ambiguity is similar to
Kaplan-Manohar ambiguity in ChPT\cite{Kaplan86}. It will be studied at
elsewhere so that we do not further discuss it here. In next
section, rigorous calculation will show that our results in this paper is
independent of $\kappa$.  

To conclude this section, the ChCQM lagrangian including the lowest vector
meson resonances and light current quark masses is
\begin{eqnarray}\label{2.13}
{\cal L}_{\chi}&=&i\bar{q}(\sla{\pa}+\sla{\Gamma}+
  g_{_A}{\slash\!\!\!\!\Delta}\gamma_5-i\sla{V})q-m\bar{q}q
  -\frac{1}{2}\bar{q}(\xi^{\dag}\td{\chi}\xi^{\dag}
  +\xi\td{\chi}^{\dag}\xi)q
  -\frac{\kappa}{2}\bar{q}(\xi^{\dag}\td{\chi}\xi^{\dag}
 -\xi\td{\chi}^{\dag}\xi)\gamma_5q \nonumber \\ 
   &&+\frac{F^2}{16}<\nabla_\mu U\nabla^\mu U^{\dag}>
   +\frac{1}{4}m_0^2<V_\mu V^{\mu}>.
\end{eqnarray}
The effective lagrangian describing interaction of vector meson resonances
will be generated via loop effects of constituent quarks.

\section{Leading Order Effective Lagrangian in $N_c^{-1}$ Expansion}

In this section, we will derive relevant effective lagrangian via
calculating one-loop diagrams of constituent quarks, which is at
the leading order in $N_c^{-1}$ expansion. In our calculation, the light
current quark masses will be treated as perturbation and be expanded to
the leading order. Pseudoscalar mesons which are localed in
external line should satisfy soft pion theorem, i.e., $k^2\rightarrow
0$(where $k^2$ denotes the four-momentum square of pseudoscalar mesons).
However, $q^2$ is the four-momentum square of vector mesons, and
obviously it is not very small comparing with chiral symmetry
spantoneously broken scale. Therefore, the higher order terms of $q^2$ in
the chiral expansion, have significant contribution which can not be
neglected. Or in the other words, the chiral expansion at vector
meson energy scale converge slowly. We will rigorously calculate all
$q^2$-dependent term contribution.

We start with constituent quark lagrangian(~\ref{2.1}), and define vector
external source $\bar{V}_\mu^a(a=0,1,\cdots,8)$, axial-vector
external source $\Delta_\mu^a$, scalar external source
$S^a$ and pseudoscalar external source $P^a$ as follows
\begin{eqnarray}\label{3.1}
\bar{V}_\mu^\alpha&=&\frac{1}{2}<\lambda^\alpha(V_\mu+i\Gamma_\mu)>,
 \hspace{0.8in}
\Delta_\mu^a=\frac{1}{2}<\lambda^a\Delta_\mu>,\nonumber \\
S^a&=&\frac{1}{4}<\lambda^a(\xi^{\dag}\td{\chi}\xi^{\dag}
  +\xi\td{\chi}^{\dag}\xi)>, \hspace{0.55in}
P^a=\frac{\kappa}{4}<\lambda^a(\xi^{\dag}\td{\chi}\xi^{\dag}
  -\xi\td{\chi}^{\dag}\xi)>
\end{eqnarray}
(where $\lambda^1,\cdots,\lambda^8$ are SU(3) Gell-Mann matrices and 
$\lambda^0=\sqrt{\frac{2}{3}}$). Then in lagrangian(~\ref{2.1}), the 
terms associating with constituent quark fields can be rewritten as follow
\begin{eqnarray}\label{3.2}
{\cal L}_\chi^q=\bar{q}(i\sla{\pa}-m)q
  +\bar{V}_\mu^a\bar{q}\lambda^a\gamma^\mu q
  +ig_A\Delta_\mu^a\bar{q}\lambda^a\gamma^\mu\gamma_5q
  -S^a\bar{q}\lambda^aq-P^a\bar{q}\lambda^a\gamma_5q.
\end{eqnarray}

The effective action describing meson interaction can be obtained via
integrating over degrees of freedom of fermions
\begin{equation}\label{3.3}
e^{iS_{\rm eff}}\equiv\int{\cal D}\bar{q}{\cal D}qe^{i\int d^4x{\cal
   L}_\chi(x)}=<vac,out|in,vac>_{\bar{V},\Delta,S,P},
\end{equation}
where $<vac,out|in,vac>_{\bar{V},\Delta,S,P}$ is vacuum expectation value
in presence external sources. The above path integral can be performed
explicitly, and heat kernal method\cite{Sch51,Ball89} has been used to
regulate the result. However, this method is extremely difficult to
compute very high order contributions in practice. In {\bf I} we have
provided an equivalent and efficient method to evaluate the effective
action via calculating one-loop diagrams of constituent quarks directly.
This method can capture all high order contributions of the chiral
expansion.

In interaction picture, the equation(~\ref{3.3}) is rewritten as follow
\begin{eqnarray}\label{3.4}
e^{iS_{\rm eff}}&=&<0|{\cal T}_qe^{i\int d^4x{\cal L}^{\rm I}_\chi(x)}|0>
       \nonumber \\ 
 &=&\sum_{n=1}^\infty i\int d^4p_1\frac{d^4p_2}{(2\pi)^4}
  \cdots\frac{d^4p_n}{(2\pi)^4}\tilde{\Pi}_n(p_1,\cdots,p_n)
  \delta^4(p_1-p_2-\cdots-p_n) \nonumber \\
&\equiv&i\Pi_1(0)+\sum_{n=2}^\infty i\int \frac{d^4p_1}{(2\pi)^4}
  \cdots\frac{d^4p_{n-1}}{(2\pi)^4}\Pi_n(p_1,\cdots,p_{n-1}),
\end{eqnarray}
where ${\cal T}_q$ is time-order product of constituent quark fields,
${\cal L}_{\chi}^{\rm I}$ is interaction part of lagrangian(~\ref{3.2}), 
$\tilde{\Pi}_n(p_1,\cdots,p_n)$ is one-loop effects of constituent quarks
with $n$ external sources, $p_1,p_2,\cdots,p_n$ are four-momentas of $n$
external sources respectively and
\begin{equation}\label{3.5}
\Pi_n(p_1,\cdots,p_{n-1})=\int d^4p_n\tilde{\Pi}_n(p_1,\cdots,p_n)
  \delta^4(p_1-p_2-\cdots-p_n).
\end{equation}
To get rid of all disconnected diagrams, we have
\begin{eqnarray}\label{3.6}
S_{\rm eff}&=&\sum_{n=1}^\infty S_n, \nonumber \\
S_1&=&\Pi_1(0),  \\
S_n&=&\int \frac{d^4p_1}{(2\pi)^4}\cdots\frac{d^4p_{n-1}}
  {(2\pi)^4}\Pi_n(p_1,\cdots,p_{n-1}), \hspace{0.8in}(n\geq 2)\nonumber.
\end{eqnarray}
Hereafter we will call $S_n$ as $n$-point effective action.

The $S_1$ is tadpole-loop contribution of constituent quarks, which is
independent of the purpose this paper. The two-point effective action 
$S_2$ has been evaluated in {\bf I},
\begin{eqnarray}\label{3.7}
S_2&=&\frac{F_0^2}{16}\int d^4x<\nabla_\mu U\nabla^\mu U^{\dag}>+
 \int \frac{d^4q}{(2\pi)^4}\Pi_2^V(q), \nonumber \\
\Pi_2^V(q)&=&-\frac{1}{2}A(q^2)(q^2\delta_{\mu\nu}-q_\mu
  q_\nu)<\bar{V}^\mu(q)\bar{V}^\nu(-q)>,
\end{eqnarray}
where kinetic term of pseudoscalar mesons has been renormalized, and
$A(q^2)$ is defined as follow
\begin{eqnarray}\label{3.8}
A(q^2)=g^2-\frac{N_c}{\pi^2}\int_0^1 dt\cdot t(1-t)\ln{(1
 -\frac{t(1-t)q^2}{m^2})}.
\end{eqnarray}
Here a universal constant of the model, $g$, is defined to absorbe
logarithmic divergence from quark loop integral
\begin{eqnarray}\label{3.9}
g^2=\frac{8}{3}\frac{N_c}{(4\pi)^{D/2}}(\frac{\mu^2}{m^2})^{\ep/2}
  \Gamma(2-\frac{D}{2}).
\end{eqnarray}
In {\bf I} we have fitted $g\equiv \pi^{-1}\sqrt{N_c/3}$ which satisfy
the first KSRF sum rule\cite{KSRF} rigorously. 

\subsection{Three-point effective action}

Up to the leading order of light current quark masses, there are three
kinds of three-point effective action. They are made up of by external
sources $\bar{V}\Delta\Delta$, $\bar{V}\bar{V}S$ and $\bar{V}\Delta P$
respectively. The effective action with external source
$\bar{V}\Delta\Delta$ has been obtained in {\bf I},
\begin{eqnarray}\label{3.10}
S_3^{(1)}=-\frac{g_A^2}{2}\int d^4x\int\frac{d^4q}{(2\pi)^4}
  e^{iq\cdot x}B(q^2)q^\mu <\bar{V}^\nu(q)[\Delta_\mu(x),\Delta_\nu(x)]>,
\end{eqnarray}
where
\begin{eqnarray}\label{3.11}
B(q^2)=-g^2&+&\frac{N_c}{2\pi^2}\int_0^1dt_1\cdot t_1\int_0^1dt_2
 (1-t_1t_2)[1+\frac{m^2}{m^2-t_1(1-t_1)(1-t_2)q^2} \nonumber \\
 &+&\ln{(1-\frac{t_1(1-t_1)(1-t_2)q^2}{m^2})}].
\end{eqnarray}

Let us now calculate three-point effective action with external source
$\bar{V}\bar{V}S$. Note that since $\bar{V}_\mu=V_\mu+ie{\cal
Q}A_\mu+i\pi\pa_\mu\pi+\cdots$, for $\omega-\rho^0$ mixing or
$\omega\pi\pi$ coupling vertices, scalar external source $S$ will reduce
to constant matrix.  
\begin{eqnarray}\label{3.12}
iS_3^{(2)}&=&\frac{i}{2}\int d^4xd^4yd^4z\cdot\bar{V}_\mu^a(x)
 \bar{V}_\nu^b(y)S^c<0|T\{\bar{q}(x)\gamma^\mu\lambda^aq(x)
 \bar{q}(y)\gamma^\nu\lambda^bq(y)\bar{q}(z)\lambda^cq(z)\}|0>
   \nonumber \\
&=&-i\int\frac{d^4q}{(2\pi)^4}\frac{d^4l}{(2\pi)^4}
  <\bar{V}_\mu(q)\bar{V}_\nu(-q)S>Tr_{(c,L)}\{S_F(l)\gamma^\nu
  S_F^2(l+q)\gamma^\mu\},
\end{eqnarray}
where $Tr_{(c,L)}$ denotes trace taking over color and Lorentz space,
$S_F(k)=i(\sla{k}-m+i\ep)^{-1}$ is propagator of constituent quark fields
in momentum space. The direct calculation will give
\begin{eqnarray}\label{3.13}
S_3^{(2)}&=&\frac{N_c}{12\pi^2m}
  \int d^4x\int\frac{d^4q}{(2\pi)^4}e^{iq\cdot x}
  h_0(q^2)(q^2\delta_{\mu\nu}-q_\mu q_\nu)
  <\bar{V}^\mu(q)\bar{V}^\nu(x)S(x)> \nonumber \\
&=&\frac{N_c}{24\pi^2m}
  \int d^4x\int\frac{d^4q}{(2\pi)^4}e^{iq\cdot x}
  h_0(q^2)(q^2\delta_{\mu\nu}-q_\mu q_\nu)
  <\bar{V}^\mu(q)\bar{V}^\nu(x)(\xi\td{\chi}^{\dag}\xi
  +\xi^{\dag}\td{\chi}\xi^{\dag})>,
\end{eqnarray}
where
\begin{eqnarray}\label{3.14}
h_0(q^2)=\int_0^1 dt\frac{6t(1-t)}{1-t(1-t)q^2/m^2}.
\end{eqnarray}

Next, we calculate three-point effective action with external
source $\bar{V}\Delta P$.
\begin{eqnarray}\label{3.15}
iS_3^{(3)}&=&-g_A\int d^4xd^4yd^4z\cdot
 <0|T\{\bar{q}(x)\gamma^\mu\lambda^aq(x)
 \bar{q}(y)\gamma^\nu\gamma_5\lambda^bq(y)
 \bar{q}(z)\gamma_5\lambda^cq(z)\}|0>
\bar{V}_\mu^a(x)\Delta_\nu^b(y)P^c(z)\nonumber \\
&=&g_A\int\frac{d^4q}{(2\pi)^4}\frac{d^4k}{(2\pi)^4}\frac{d^4l}
  {(2\pi)^4}\{<\bar{V}_\mu(q)\Delta_\nu(k)P(-q-k)>Tr_{(c,L)}[S_F(l)
  \gamma^\nu\gamma_5S_F(l-k)\gamma_5S_F(l+q)\gamma^\mu] \nonumber \\
&&\hspace{0.5in}
+<\bar{V}_\mu(q)P(-q-k)\Delta_\nu(k)>Tr_{(c,L)}[S_F(l)\gamma^\mu
  S_F(l-q)\gamma_5S_F(l+k)\gamma^\mu\gamma_5]\}.
\end{eqnarray}
Due to $\Delta_\mu\propto\pa_\mu\pi+\cdots$ and $P\propto\pi+\cdots$, for
purpose of this paper, the soft pion theroem tell us $k^2\rightarrow 0$
and $(k+q)^2\rightarrow 0$. In addition, we can find that
$k_\mu\Delta^\mu(k)\Rightarrow k^2\pi(k)\rightarrow 0$. Then performing 
the loop-integral in eq.(~\ref{3.15}), and employing the above
discussion in our calculation, we obtain
\begin{eqnarray}\label{3.16}
S_3^{(3)}&=&-\frac{N_cm}{4\pi^2}g_A\int\frac{d^4q}{(2\pi)^4}
 \frac{d^4k}{(2\pi)^4}\{\alpha_1(q^2)<[\bar{V}_\mu(q),\Delta^\mu(k)]
 P(-k-q)>\nonumber \\
&&\hspace{0.5in}+\alpha_2(q^2)(q^2\delta_{\mu\nu}-q_\mu q_\nu)
 <[\bar{V}^\nu(q),\Delta^\nu(k)]P(-k-q)>\},  
\end{eqnarray}
where
\begin{eqnarray}\label{3.17}
\alpha_1(q^2)&=&(\frac{4\pi\mu^2}{m^2})^{\ep/2}\Gamma(2-\frac{D}{2})
 -\int_0^1dt_1\int_0^1dt_2\{2t_1\ln{(1-\frac{t_1(1-t_1)(1-t_2)q^2}{m^2})}
    \nonumber \\ &&\hspace{1.4in}
  +\frac{t_1t_2(1-t_1)q^2}{m^2-t_1(1-t_1)(1-t_2)q^2}\}, \nonumber \\
\alpha_2(q^2)&=&\int_0^1dt_1\int_0^1dt_2\frac{2t_1(1-t_1)}
  {m^2-t_1(1-t_1)(1-t_2)q^2}.
\end{eqnarray}

The third three-point effective action $S_3^{(3)}$ is $O(m_q)$ and
free parameter $\kappa$-dependent. However, if $\bar{V}_\mu=\omega_\mu$,
$S_3^{(3)}$ vanish, and if $\bar{V}_\mu=\rho_\mu^0\lambda^3$, $S_3^{(3)}$
provide an isospin conservation $\rho\pi\pi$ vertex which is order
$m_u+m_d$ and much smaller that leading order vertex. Thus the
contribution from $S_3^{(3)}$ will be omitted in this paper. 

\subsection{Four-point effective action}

There is only one four-point effective action relating to
$\omega\rightarrow\pi^+\pi^-$ decay. It is made up of by four external
source $\bar{V}\Delta\Delta S$. As shown in the above subsection, here $S$
reduces to a constant matrix.
\begin{eqnarray}\label{3.18}
iS_4&=&\frac{g_A^2}{2}\int d^4xd^4yd^4zd^4w\cdot\bar{V}_\mu^a(x)
  \Delta_\nu^b(y)\Delta_\sigma^c(z)S^d(w) \nonumber \\
&&\hspace{0.5in}\times<0|T\{\bar{q}(x)\gamma^\mu\lambda^aq(x)
  \bar{q}(y)\gamma^\nu\gamma_5\lambda^bq(y)
\bar{q}(z)\gamma^\sigma\gamma_5\lambda^cq(z)\bar{q}(w)\lambda^dq(w)\}|0>
  \nonumber \\
&=&g_A^2\int\frac{d^4q}{(2\pi)^4}\frac{d^4k}{(2\pi)^4}\frac{d^4l}{(2\pi)^4}  
  \nonumber \\
&&\times\{<\bar{V}_\mu(q)\Delta_\nu(k)\Delta_\sigma(-k-q)S>
  Tr_{(c.L)}[S_F^2(l)\gamma^\mu S_F(l-q)\gamma^\nu\gamma_5
  S_F(l-q-k)\gamma^\sigma\gamma_5] \nonumber \\
&&+\;<\bar{V}_\mu(q)\Delta_\nu(k)S\Delta_\sigma(-k-q)>
  Tr_{(c.L)}[S_F(l+q+k)\gamma^\mu S_F(l+k)\gamma^\nu\gamma_5
  S_F^2(l)\gamma^\sigma\gamma_5] \nonumber \\
&&+\;<\bar{V}_\mu(q)S\Delta_\nu(k)\Delta_\sigma(-k-q)>
  Tr_{(c.L)}[S_F(l+q)\gamma^\mu S_F^2(l)\gamma^\nu\gamma_5
  S_F(l-k)\gamma^\sigma\gamma_5]\}.
\end{eqnarray}
To perform integral of four-momenta $l$ in the above equation and employ
indentities in Appendix to simplify result, we can obtain
\begin{eqnarray}\label{3.19}
S_4&=&-\frac{N_c}{8\pi^2m}g_A^2\int d^4x\int\frac{d^4q}{(2\pi)^4}
 e^{iq\cdot x}(\delta_{\mu\nu}q_\sigma-\delta_{\mu\sigma}q_\nu)
  \nonumber \\ &&\times
  \{h_1(q^2)<\{\bar{V}^\mu(q),S(x)\}\Delta^\nu(x)\Delta^\sigma(x)>
   +h_2(q^2)<\bar{V}^\mu(q)\Delta^\nu(x)S(x)\Delta^\sigma(x)>\},
\end{eqnarray}
where
\begin{eqnarray}\label{3.20}
h_1(q^2)&=&\int_0^1dt_1\cdot t_1^2\int_0^1dt_2(1-t_2)
  \frac{3-2t_1^2t_2(1+2t_1)(1-t_2)q^2/m^2}{[1-t_1^2t_2(1-t_2)q^2/m^2]^2},
   \nonumber \\
h_2(q^2)&=&\int_0^1dt_1\cdot t_1^2\int_0^1dt_2(1-t_2)
  \frac{4(1-t_1)[3-4t_1^2t_2(1-t_2)q^2/m^2]}
   {[1-t_1^2t_2(1-t_2)q^2/m^2]^2}.
\end{eqnarray}

\subsection{Relevant effective vertices at tree level}

In the following we will give all relevant effective vertice at tree
level. Since one-loop correction of pseudoscalar mesons will be
calculated, we also need to include four-pseudoscalar meson vertices,
which will be derived from $O(p^4)$ effective lagrangian of this formlism.
The effective vertices involving electromagnetic interaction have been
calculated up to one-loop level in {\bf I}. We will quote them in sect. 5
directly. All effective vertices can be divided into two part: one is
isospin conservation and anthor is isospin broken. In addition, we should
point out that, so far, all meson fields are still non-physical.
The physical meson fields can be obtained via the following field
rescaling which make kinetic terms of pseudoscalar mesons and vector
mesons into standard form
\begin{eqnarray}\label{rescale}
&&\rho_\mu^0\lraw\frac{1}{g}\rho_\mu^0,\hspace{1in}
  \omega_\mu\lraw\frac{1}{g}\omega_\mu, \nonumber \\
&&\pi\lraw\frac{2}{f_\pi}\pi,\hspace{1in}
  K\lraw\frac{2}{f_\pi}K.
\end{eqnarray}
Since in this paper $K$-mesons only appear as intermediate states in
one-loop diagrams, for sake of convenience, we neglect the difference
between $f_\pi$ and $f_K$(since the results yielded by this difference are
twofold suppressed by light current quark mass expansion and $N_c^{-1}$
expansion).

The $\rho^0-\omega$ mixing vertex, which breaks isospin symmetry, is
included in eq.(~\ref{3.13})
\begin{eqnarray}\label{3.21}
{\cal L}_{\omega\rho}&=&\frac{N_c}{12\pi^2g^2m}\int\frac{d^4q}{(2\pi)^4}
  e^{iq\cdot x}h_0(q^2)(q^2\delta_{\mu\nu}-q_\mu q_\nu)\omega^\mu(q)
 \rho^{0\nu}(x)
 <\lambda^3(\xi{\cal M}\xi+\xi^{\dag}{\cal M}\xi^{\dag})> \\
&=&\frac{N_c}{6\pi^2g^2}\frac{m_u-m_d}{m}
 \int\frac{d^4q}{(2\pi)^4}e^{iq\cdot x}h_0(q^2)
  (q^2\delta_{\mu\nu}-q_\mu q_\nu)\omega^\mu(q)\rho^{0\nu}(x)+\cdots.
\end{eqnarray}

The isospin symmetry unbroken vector$\rightarrow\vphi\vphi$ vertex is
included in eqs.(~\ref{3.7}) and (~\ref{3.10})
\begin{eqnarray}\label{3.22}
{\cal L}_{V\vphi\vphi}^{(\Delta I=0)}
=-2\int\frac{d^4q}{(2\pi)^4}e^{iq\cdot x}b(q^2)q_\mu
  <V_\nu(q)[\Delta^\mu,\Delta^\nu]>,
\end{eqnarray}
where
\begin{equation}\label{3.23}
b(q^2)=\frac{1}{gf_\pi^2}[A(q^2)+g_A^2B(q^2)].
\end{equation}

The isospin symmetry broken vector$\rightarrow\vphi\vphi$ vertex is
include in eqs.(~\ref{3.13}) and (~\ref{3.19})
\begin{eqnarray}\label{3.24}
{\cal L}_{V\vphi\vphi}^{(\Delta I=1)}
&=&-\frac{N_c}{12\pi^2mg}\int\frac{d^4q}{(2\pi)^4}e^{iq\cdot x}
  (\delta_{\mu\nu}q_\sigma-\delta_{\mu\sigma}q_\nu) \nonumber \\
&&\times\{[h_0(q^2)+\frac{3}{4}g_A^2h_1(q^2)]
<\{\bar{V}^\mu(q),\xi{\cal M}\xi+\xi^{\dag}{\cal M}\xi^{\dag}\}
 \Delta^\nu(x)\Delta^\sigma(x)> \nonumber \\
&&\;-\frac{3}{4}g_A^2h_2(q^2)
 <\bar{V}^\mu(q)\Delta^\nu(x)(\xi{\cal M}\xi
 +\xi^{\dag}{\cal M}\xi^{\dag})\Delta^\sigma(x)>\}. 
\end{eqnarray}

In particular, define
\begin{equation}\label{3.25}
s(q^2)=\frac{4}{gf_\pi^2}
[h_0(q^2)+\frac{3}{4}g_A^2(h_1(q^2)-\frac{h_2(q^2)}{2})],
\end{equation}
we have
\begin{eqnarray}\label{3.26}
{\cal L}_{\rho^0\pi\pi}&=&i\int\frac{d^4q}{(2\pi)^4}
  e^{iq\cdot x}b(q^2)(q^2\delta_{\mu\nu}-q_\mu q_\nu)
  \rho^{0\mu}(q)[\pi^+(x)\pa^\nu\pi^-(x)-\pa^\nu\pi^+(x)\pi^-(x)]
        \nonumber \\
{\cal L}_{\omega\pi\pi}&=&-\frac{iN_c}{12\pi^2}\frac{m_u-m_d}{m}
  \int\frac{d^4q}{(2\pi)^4}e^{iq\cdot x}s(q^2)
 (q^2\delta_{\mu\nu}-q_\mu q_\nu)\omega^\mu(q) \nonumber \\
 &&\hspace{1in}\times
 [\pi^+(x)\pa^\nu\pi^-(x)-\pa^\nu\pi^+(x)\pi^-(x)] \nonumber \\
{\cal L}_{\rho^0KK}^{(\Delta I=0)}&=&\frac{i}{2}
 \int\frac{d^4q}{(2\pi)^4}e^{iq\cdot x}b(q^2)
 (q^2\delta_{\mu\nu}-q_\mu q_\nu)\rho^{0\mu}(q) \nonumber \\
&\times&
 \{[K^+(x)\pa^\nu K^-(x)-\pa^\nu K^+(x)K^-(x)]
  -[K^0(x)\pa^\nu\bar{K}^0(x)-\pa^\nu K^0(x)\bar{K}^0(x)]\}\nonumber\\
{\cal L}_{\rho^0KK}^{(\Delta I=1)}&=&
 -\frac{iN_c}{24\pi^2}\frac{m_u-m_d}{m}
 \int\frac{d^4q}{(2\pi)^4}e^{iq\cdot x}s(q^2)
 (q^2\delta_{\mu\nu}-q_\mu q_\nu)\rho^{0\mu}(q) \\
 &\times&
 \{[K^+(x)\pa^\nu K^-(x)-\pa^\nu K^+(x)K^-(x)]
  +[K^0(x)\pa^\nu\bar{K}^0(x)-\pa^\nu K^0(x)\bar{K}^0(x)]\}\nonumber\\
{\cal L}_{\omega KK}^{(\Delta I=0)}&=&\frac{i}{2}
 \int\frac{d^4q}{(2\pi)^4}e^{iq\cdot x}b(q^2)
 (q^2\delta_{\mu\nu}-q_\mu q_\nu)\omega^{\mu}(q) \nonumber \\
&\times&
 \{[K^+(x)\pa^\nu K^-(x)-\pa^\nu K^+(x)K^-(x)]
  +[K^0(x)\pa^\nu\bar{K}^0(x)-\pa^\nu K^0(x)\bar{K}^0(x)]\}\nonumber\\  
{\cal L}_{\omega KK}^{(\Delta I=1)}&=&
 -\frac{iN_c}{24\pi^2}\frac{m_u-m_d}{m}
 \int\frac{d^4q}{(2\pi)^4}e^{iq\cdot x}s(q^2)
 (q^2\delta_{\mu\nu}-q_\mu q_\nu)\omega^{\mu}(q) \nonumber \\
&\times&
 \{[K^+(x)\pa^\nu K^-(x)-\pa^\nu K^+(x)K^-(x)]
  -[K^0(x)\pa^\nu\bar{K}^0(x)-\pa^\nu K^0(x)\bar{K}^0(x)]\}\nonumber
\end{eqnarray}

Up to $O(p^4)$, the tree level four-pseudoscalar meson effective
lagrangian has been derived in {\bf I}. The isospin symmetry unbroken
four-pseudoscalar vertex is included in the following lagrangian
\begin{eqnarray}\label{3.27}
{\cal L}_{4\vphi}^{(\Delta I=0)}&=&
\frac{f_\pi^2}{16}<\na_\mu U\na^\mu U^{\dag}>
+\frac{N_c}{12(4\pi)^2}<\na_\mu U\na_\nu U^{\dag}\na_\mu U\na_\nu
U^{\dag}> \nonumber \\
&&-\frac{N_c}{12(4\pi)^2}(1-g_A^4)
<\na_\mu U\na^\mu U^{\dag}\na_\nu U\na^\nu U^{\dag}>,
\end{eqnarray}
where we have used $g^2=N_c/(3\pi^2)$. The isospin symmetry broken
four-pseudoscalar vertex is proportional to $m_d-m_u$, which is included
in the following lagrangian
\begin{eqnarray}\label{3.28}
{\cal L}_{4\vphi}^{(\Delta I=1)}=
\frac{f_\pi^2}{8}B_0<{\cal M}(U+U^{\dag})>
+\frac{N_cm}{(4\pi)^2}g_A^2<\na_\mu U\na^\mu U^{\dag}
 ({\cal M}U^{\dag}+U{\cal M})>.
\end{eqnarray}

I can be found that the eqs.(~\ref{3.21})-(~\ref{3.28}) are free
parameter $\kappa$ independent. Moreover, we can see that every
vector$\rightarrow\vphi\vphi$ in eq.(~\ref{3.26}) includes an antisymmetry
factor $(q^2\delta_{\mu\nu}-q_\mu q_\nu)$(where $q_\mu$ denotes
four-momenta of vector mesons). Thus the first term of eq.(~\ref{3.28})
does not contribute to $\omega\rightarrow\pi^+\pi^-$ decay via
pseudoscalar meson loops. This antisymmetry factor also constrains that 
the vertices with one of factors $K\bar{K}$, $\pa_\mu K\pa^\mu\bar{K}$ and
$K\pa^2\bar{K}$ do not contribute to $\omega\rightarrow\pi^+\pi^-$ decay
via pseudoscalar meson loops. Then the relevant four-pseudoscalar vertices
can explicitly read as follows,
\begin{eqnarray}\label{3.29}
{\cal L}_{4\pi}&=&\frac{2}{f_\pi^2}(\pi^+\pa_\mu\pi^-)
 (\pi^+\pa^\mu\pi^-)+\frac{2N_c}{3\pi^2f_\pi^4}\pa_\mu\pi^+\pa_\nu\pi^- 
   [\pa^\mu\pi^+\pa^\nu\pi^--(1-g_A^4)\pa^\mu\pi^-\pa^\nu\pi^+],
       \nonumber \\
{\cal L}_{KK\pi\pi}^{(\Delta I=0)}&=&
  \frac{4}{f_\pi^2}(K^+\pa_\mu K^-+\pa_\mu K^0\bar{K}^0)
  \pi^+\pa^\mu\pi^- \nonumber \\ 
&&+\frac{2N_c}{3\pi^2f_\pi^4}(\pa_\mu K^+\pa_\nu K^-
  +\pa_\nu K^0\pa_\mu\bar{K}^0)[2\pa^\mu\pi^+\pa^\nu\pi^-
  -(1-g_A^4)\pa^\mu\pi^-\pa^\nu\pi^+], \nonumber \\
{\cal L}_{KK\pi\pi}^{(\Delta I=1)}&=&\frac{16N_c}{\pi^2f_\pi^4}g_A^2m
(m_uK^+\pa_\mu K^--m_dK^0\pa_\mu\bar{K}^0)\pi^+\pa^\mu\pi^-.
\end{eqnarray}

\section{One-loop Corrections of Pseudoscalar Mesons}

In this section we calculate one-loop correction of mesons. Because of
$m_{_V}^2>m_{_K}^2>>m_\pi^2$, it can be expected that the dominant
contribution comes from one-loop digrams of pseudoscalar mesons. In
addition, we can treat pion as massless particle but must take
$m_{_K}^2\neq 0$. This difference is very important, since $\pi$-loop
yields imaginary part of ${\cal T}$-matrix but $K$-loop does not at
$m_{\omega}$ scale. In our calculation, the mass difference between
$K^\pm$ and $K^0$ is also neglected.

There are three kinds of one-loop
diagrams correcting to ``direct'' $\omega\pi\pi$ couping and
$\omega-\rho^0$ mixing(fig. 1 and fig. 2).

\begin{figure}[hptb]
   \centerline{
   \psfig{figure=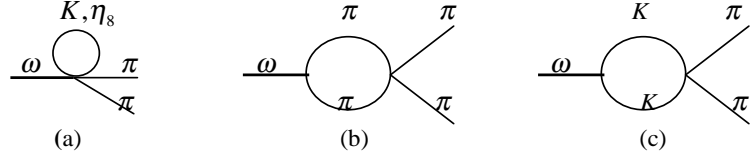,width=4.5in}}
\begin{minipage}{5in}
   \caption{One-loop correction to ``direct'' $\omega\pi\pi$ couping.}
\end{minipage}
\end{figure}
  
\begin{figure}[hptb]
   \centerline{
   \psfig{figure=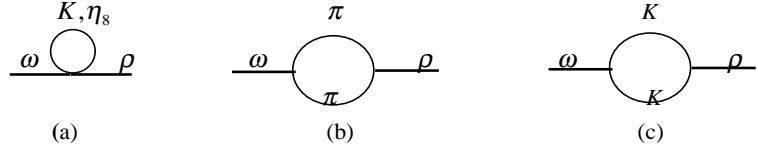,width=4.5in}}
\begin{minipage}{5in}
   \caption{One-loop correction to $\omega-rho^0$ mixing.}
\end{minipage}    
\end{figure}  

It must be pointed out that, in ${\cal T}$-matrices yielded by figure
1-(b) and figure 2-(b), the contribution of imaginary part is dominant. We
have shown in {\bf I} that it can not ensure unitarity of $S$-matrix if we
only calculate figure 1-(b) and figure 2-(b). The
unitarity can be ensured through summing over all diagrams in chain
approximation(fig. 3 and fig. 4).

\begin{figure}[hptb]
   \centerline{
   \psfig{figure=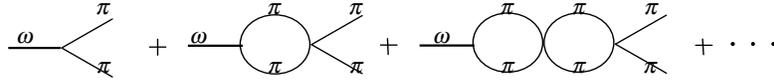,width=4.5in}}
\begin{minipage}{5in}
   \caption{Chain approximation for ``direct'' $\omega\pi\pi$ couping.}
\end{minipage}
\end{figure}

\begin{figure}[hptb]
   \centerline{
   \psfig{figure=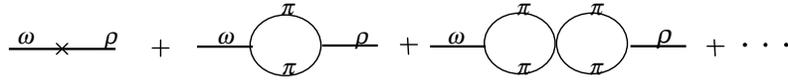,width=4.5in}}
\begin{minipage}{5in}
   \caption{Chain approximation for $\omega-\rho^0$ mixing.}
\end{minipage}
\end{figure}

\subsection{Tadpole diagram}

Since pion is treated as massless particle, the nonzero tadpole diagram
contribution is yielded by $K$ or $\eta_8$ mesons(fig. 1-(a) and fig.
2-(a)). For sake of convenience, here we assume $m_{\eta_8}=m_{_K}$.

\begin{description}
\item[A] Correction to $\omega-\rho^0$ mixing
\end{description}
The tree level $\omega\rho^0\vphi\vphi$ is contained in eq.(35),
\begin{eqnarray}\label{4.1}
{\cal L}_{\omega\rho^0\vphi\vphi}=
 -\frac{N_c}{12\pi^2g^2f_\pi^2m}\int\frac{d^4q}{(2\pi)^4}
  e^{iq\cdot x}h_0(q^2)(q^2\delta_{\mu\nu}-q_\mu q_\nu)\omega^\mu(q)
 \rho^{0\nu}(x)
 <\lambda^3(\vphi^2{\cal M}+{\cal M}\vphi^2+\vphi{\cal M}\vphi)>.  
\end{eqnarray}
In momentum space, the calculation on fig. 2-(a) is straighforward
\begin{eqnarray}\label{4.2}
{\cal L}_{\omega\rho^0}^{(tad)}&=&
 -\frac{N_c}{12\pi^2g^2f_\pi^2m}\int\frac{d^4q}{(2\pi)^4}
  e^{iq\cdot x}h_0(q^2)(q^2\delta_{\mu\nu}-q_\mu q_\nu)\omega^\mu(q)
 \rho^{0\nu}(x) \nonumber \\&&\times\int\frac{d^4k}{(2\pi)^4}\frac{i}
  {k^2-m_{_K}^2+i\ep} \sum_{a=4}^{8}
  <\lambda^3(\lambda^a\lambda^a{\cal M}+{\cal M}\lambda^a\lambda^a
  +\lambda^a{\cal M}\lambda^a)>.
\end{eqnarray}
The generators $\lambda^a$ of SU(N) obey the completeness relations
\begin{eqnarray}\label{4.3}
\sum_{a=1}^{N^2-1}<\lambda^aA\lambda^aB>&=&-\frac{2}{N}<AB>+2<A><B>,
  \nonumber \\
\sum_{a=1}^{N^2-1}<\lambda^aA><\lambda^aB>&=&\;
  2<AB>-\frac{2}{N}<A><B>.
\end{eqnarray}
Then we have
\begin{eqnarray}\label{4.4}
\sum_{a=4}^{8}<\lambda^3(\lambda^a\lambda^a{\cal M}+{\cal 
 M}\lambda^a\lambda^a+\lambda^a{\cal M}\lambda^a)>
=\frac{16}{3}<\lambda^3{\cal M}>=\frac{16}{3}(m_u-m_d).
\end{eqnarray}
Substituting eq.(~\ref{4.4}) into eq.(~\ref{4.2}) and performing
loop integral, we obtain
\begin{eqnarray}\label{4.5}
{\cal L}_{\omega\rho^0}^{(tad)}&=&-\frac{4}{3}\zeta
 \frac{N_c}{6\pi^2g^2}\frac{m_u-m_d}{m}\int\frac{d^4q}{(2\pi)^4} 
  e^{iq\cdot x}h_0(q^2)(q^2\delta_{\mu\nu}-q_\mu q_\nu)\omega^\mu(q)
 \rho^{0\nu}(x),
\end{eqnarray}
where
\begin{equation}\label{4.6}
\zeta=\frac{2\lambda}{(4\pi)^2}\frac{m_{_K}^2}{f_\pi^2},\hspace{1in}
\lambda=(\frac{4\pi\mu^2}{m_{_K}^2})^{\ep/2}\Gamma(1-\frac{D}{2}).
\end{equation}
Here we define a constant $\lambda$ to absorbe quadratic divergence from
meson loop integral. Its value has been determined as $\lambda=2/3$ in
{\bf I} by OZI rule. 

\begin{description}
\item[B] Correction to ``direct'' $\omega\pi\pi$ mixing
\end{description}

The isospin symmetry broken $\omega$-4$\vphi$ couping vertex is included
in eq.(~\ref{3.24}). Expanding eq.(~\ref{3.24}) to contain four
pseudoscalar meson fields, we can obtain
\begin{eqnarray}\label{4.7}
{\cal L}_{\omega\pi\pi}^{(tad)}&=&-\frac{N_c}{12\pi^2f_\pi^2m}
 \int\frac{d^4q}{(2\pi)^4}e^{iq\cdot x}s(q^2)(\delta_{\mu\nu}q_\sigma
 -\delta_{\mu\sigma}q_\nu)\omega^\mu(q)\int\frac{d^4k}{(2\pi)^4}\frac{i}
  {k^2-m_{_K}^2+i\ep} \nonumber \\&&\times\sum_{a=4}^8
 \{<I_2{\cal M}(\pa^\nu\pi\lambda^a[\lambda^a,\pa^\sigma\pi]
  +\lambda^a[\lambda^a,\pa^\nu\pi]\pa^\sigma\pi)>\nonumber \\&&
  \;\;\;\;
 +\frac{1}{2}<I_2\pa^\nu\pi\pa^\sigma\pi
 (\lambda^a\lambda^a{\cal M}+{\cal
 M}\lambda^a\lambda^a+\lambda^a{\cal M}\lambda^a)>\}\nonumber \\
&=&\frac{10}{3}\zeta\frac{iN_c}{12\pi^2}\frac{m_u-m_d}{m}
  \int\frac{d^4q}{(2\pi)^4}e^{iq\cdot x}s(q^2)
 (q^2\delta_{\mu\nu}-q_\mu q_\nu)\omega^\mu(q) \nonumber \\ 
 &&\hspace{1in}\times
 [\pi^+(x)\pa^\nu\pi^-(x)-\pa^\nu\pi^+(x)\pi^-(x)],
\end{eqnarray}
where $I_2={\rm diag}\{1,1,0\}$, $s(q^2)$, to see eq.(~\ref{3.25}) and
eq.(~\ref{4.3}) has been used.

\subsection{$K$-loop contribution}

Here $K$-loop denotes that one-loop diagrams in fig.1-(c) and fig.2-(c).
In this subsection, $T$ will denote time-order product of $K$-meson field.
\begin{description}
\item[A] Correction to $\omega-\rho^0$ mixing
\end{description}
The $\omega-\rho^0$ effective action yielded by $K$-loop is follow
\begin{eqnarray}\label{4.8}
iS_{\omega\rho}^{(K-loop)}&=&-\int d^4xd^4y
<0|T\{{\cal L}_{\rho KK}^{(\Delta I=0)}(x)
  {\cal L}_{\omega KK}^{(\Delta I=1)}(y)+
  {\cal L}_{\rho KK}^{(\Delta I=1)}(x)
  {\cal L}_{\omega KK}^{(\Delta I=0)}(y)\}|0> \nonumber \\
&=&\frac{N_c}{3\pi^2}\frac{m_u-m_d}{m}\int\frac{d^4q}{(2\pi)^4}
 \frac{d^4l}{(2\pi)^4}b(q^2)s(q^2)(q^2\delta_{\mu\nu}
  -q_\mu q_\nu)(q^2\delta_{\alpha\beta}-q_{\alpha}q_{\beta})
  \nonumber \\&&\hspace{1in}\times \omega^\mu(q)\rho^{0\alpha}(-q)
 (l+q)^\nu l^{\beta}\Delta_{K}(l)\Delta_{K}(l+q),
\end{eqnarray}
where $\Delta_{K}(l)=i(l^2-m_{_K}^2+i\ep)^{-1}$ is propagator of
$K$-meson. Integrating over $l_\mu$ in the above equation and defining
\begin{eqnarray}\label{4.9}
\Sigma_{K}(q^2)=\frac{1}{(4\pi)^2}\{\lambda(m_{_K}^2-\frac{q^2}{6})
  +\int_0^1dt[m_{_K}^2-t(1-t)q^2]\ln{(1-\frac{t(1-t)q^2}{m_{_K}^2})}\},
\end{eqnarray}
we have
\begin{eqnarray}\label{4.10}
S_{\omega\rho}^{(K-loop)}=\frac{N_c}{6\pi^2}\frac{m_u-m_d}{m}
 \int\frac{d^4q}{(2\pi)^4}b(q^2)s(q^2)\Sigma_K(q^2)q^2
 (q^2\delta_{\mu\nu}-q_\mu q_\nu)\omega^\mu(q)\rho^{0\nu}(-q).
\end{eqnarray}
The corresponding effective lagrangian read
\begin{eqnarray}\label{4.11}
{\cal L}_{\omega\rho}^{(K-loop)}=\frac{N_c}{6\pi^2}
  \frac{m_u-m_d}{m}\int\frac{d^4q}{(2\pi)^4}e^{iq\cdot x}
  b(q^2)s(q^2)\Sigma_K(q^2)q^2
 (q^2\delta_{\mu\nu}-q_\mu q_\nu)\omega^\mu(q)\rho^{0\nu}(x).
\end{eqnarray}

\begin{description}
\item[B] Correction to ``direct'' $\omega\pi\pi$ couping
\end{description}
The ``direct'' $\omega\pi\pi$ couping effective action yielded by $K$-loop
can be evaluated as follow
\begin{eqnarray}\label{4.12}
iS_{\omega\pi\pi}^{(K-loop)}&=&-\int d^4xd^4y
  <0|T\{{\cal L}_{\rho KK}^{(\Delta I=0)}(x)
  {\cal L}_{\pi\pi KK}^{(\Delta I=1)}(y)+
  {\cal L}_{\rho KK}^{(\Delta I=1)}(x)
  {\cal L}_{\pi\pi KK}^{(\Delta I=0)}(y)\}|0> \nonumber \\
&=&-\frac{N_c}{\pi^2f_\pi^2}\frac{m_u-m_d}{m}
 \int\frac{d^4q}{(2\pi)^4}\frac{d^4k}{(2\pi)^4}\frac{d^4l}{(2\pi)^4}
 [16m^2f_\pi^{-2}b(q^2)
  -\frac{2}{3}s(q^2)-\frac{(3-g_A^4)N_c}{18\pi^2f_\pi^2}
 q^2s(q^2)] \nonumber \\&&\hspace{0.5in}\times
 (q^2\delta_{\mu\nu}-q_\mu q_\nu)(k\cdot l)l^\nu
 \omega^\mu(q)\pi^+(-q-k)\pi^-(k)\Delta_K(l)\Delta_K(l+q) \nonumber \\
&\simeq&-\frac{N_c}{\pi^2f_\pi^2}\frac{m_u-m_d}{m}
 i\int\frac{d^4q}{(2\pi)^4}\frac{d^4k}{(2\pi)^4}
 [8m^2f_\pi^{-2}b(q^2)-\frac{1}{3}s(q^2)(1+\frac{q^2N_c}{4\pi^2f_\pi^2})]
 \Sigma_K(q^2)\nonumber \\&&\hspace{0.5in}\times
  (q^2\delta_{\mu\nu}-q_{\mu}q_{\nu})k^\nu\omega^\mu(q)
  \pi^+(-q-k)\pi^-(k),
\end{eqnarray}
where we have taken soft pion limit and $3-g_A^4\simeq 3$ due to
$g_A^4\ll 0.3$. The corresponding effective lagrangian reads
\begin{eqnarray}\label{4.13}
{\cal L}_{\omega\pi\pi}^{(K-loop)}&=&\frac{iN_c}{2\pi^2f_\pi^2}
  \frac{m_u-m_d}{m}\int\frac{d^4q}{(2\pi)^4}e^{iq\cdot x}
 [8m^2f_\pi^{-2}b(q^2)-\frac{1}{3}s(q^2)(1+\frac{q^2N_c}{4\pi^2f_\pi^2})]
  \Sigma_K(q^2)\nonumber \\&&\hspace{0.5in}\times
  (q^2\delta_{\mu\nu}-q_{\mu}q_{\nu})\omega^\mu(q)
  [\pi^+(x)\pa^\nu\pi^-(x)-\pa^\nu\pi^+(x)\pi^-(x)].
\end{eqnarray}

\subsection{Chain contribution of $\pi$-loop}

Finally, we calculate chain approximation corrections of $\pi$-loop in
fig. 3 and fig. 4.

\begin{description}
\item[A] Correction to ``direct'' $\omega\pi\pi$ coupling
\end{description}
The effective action yielded by $\pi$-loop in fig. 1-(b) is evaluated as
follow
\begin{eqnarray}\label{4.14}
iS_{\omega\pi\pi}^{(\pi-loop)}&=&-\int d^4xd^4y
 <0|T\{{\cal L}_{\omega\pi\pi}(x){\cal L}_{4\pi}(y)\}|0>\nonumber \\
&=&\frac{4N_c}{3\pi^2f_\pi^2}\frac{m_u-m_d}{m}\int\frac{d^4q}{(2\pi)^4}
 \frac{d^4k}{(2\pi)^4}\frac{d^4l}{(2\pi)^4}s(q^2)
 (1+\frac{q^2N_c}{4\pi^2f_\pi^2})(q^2\delta_{\mu\nu}-q_\mu q_\nu)
   \nonumber \\&&\hspace{0.5in}\times (l\cdot k)l^\nu 
 \omega^\mu(q)\pi^+(-q-k)\pi^-(k)\Delta_{\pi}(l)\Delta_{\pi}(l+k),
\end{eqnarray}
where we have employed soft pion limit and $g_A^4\ll 3$, and
$\Delta_{\pi}(l)=i(l^2+i\ep)^{-1}$ is propagator of pion.
Integrating over $l$ in the above equation and defining
\begin{eqnarray}\label{4.15}
\Sigma_\pi(q^2)=\frac{q^2}{(4\pi)^2}\{\frac{\lambda}{6}
  +\int_0^1dt\cdot t(1-t)\ln{\frac{t(1-t)q^2}{m_{_K}^2}}
  +\frac{i}{6}Arg(-1)\theta(q^2-4m_\pi^2)\},
\end{eqnarray}
we have
\begin{eqnarray}\label{4.16}
S_{\omega\pi\pi}^{(\pi-loop)}&=&-\frac{2N_c}{3\pi^2f_\pi^2}
\frac{m_u-m_d}{m}\int\frac{d^4q}{(2\pi)^4}
\frac{d^4k}{(2\pi)^4}s(q^2)(1+\frac{q^2N_c}{4\pi^2f_\pi^2})
\Sigma_\pi(q^2)\nonumber \\&&\hspace{1in}\times
(q^2\delta_{\mu\nu}-q_\mu q_\nu)k^\nu
\omega^\mu(q)\pi^+(-q-k)\pi^-(k).
\end{eqnarray}
In eq.(~\ref{4.15}), $Arg(-1)=-\pi$ has been fitted in {\bf I} due to
requirement of unitarity. The corresponding effective lagrangian reads
\begin{eqnarray}\label{4.17}
{\cal L}_{\omega\pi\pi}^{(\pi-loop)}&=&\frac{iN_c}{3\pi^2f_\pi^2}
  \frac{m_u-m_d}{m}\int\frac{d^4q}{(2\pi)^4}e^{iq\cdot x}
  s(q^2)(1+\frac{q^2N_c}{4\pi^2f_\pi^2})\Sigma_\pi(q^2)
      \nonumber \\&&\times
  (q^2\delta_{\mu\nu}-q_{\mu}q_{\nu})\omega^\mu(q)
  [\pi^+(x)\pa^\nu\pi^-(x)-\pa^\nu\pi^+(x)\pi^-(x)].
\end{eqnarray}
Comparing eq.(~\ref{4.17}) and tree level vertex(~\ref{3.26}), we can find
that every one-loop in fig.3 contributes a factor
\begin{eqnarray}\label{4.18}
-\Xi(q^2)=-4f_\pi^{-2}(1+\frac{q^2N_c}{4\pi^2f_\pi^2})\Sigma_\pi(q^2).
\end{eqnarray}
Thus summing over all diagrams in fig. 3, we obtain
\begin{eqnarray}\label{4.19}
{\cal L}_{\omega\pi\pi}^{*}=\frac{iN_c}{12\pi^2}
  \frac{m_u-m_d}{m}\int\frac{d^4q}{(2\pi)^4}e^{iq\cdot x}
  \frac{s(q^2)}{1+\Xi(q^2)}(q^2\delta_{\mu\nu}-q_{\mu}q_{\nu})
 \omega^\mu(q)[\pi^+(x)\pa^\nu\pi^-(x)-\pa^\nu\pi^+(x)\pi^-(x)].
\end{eqnarray}

\begin{description}
\item[B] Correction to $\omega-\rho^0$ mixing
\end{description}

If in fig. 2-(b), tree level vertex ${\cal L}_{\omega\pi\pi}$ is replaced
by ${\cal L}_{\omega\pi\pi}^*$ which contains all diagram contribution in
fig. 3, then summing tree diagram and fig. 2-(b) is just chain
approximation correction to $\omega-\rho^0$ mixing. The effective action
yielded by $\pi$-loop in fig. 2-(b) is evaluated as follow
\begin{eqnarray}\label{4.20}
iS_{\omega\rho}^{(\pi-loop)}&=&-\int d^4xd^4y
<0|T\{{\cal L}_{\omega\pi\pi}^*(x){\cal L}_{\rho\pi\pi}(y)\}|0>
 \nonumber \\
&=&-\frac{N_c}{6\pi^2}\frac{m_u-m_d}{m}i\int\frac{d^4q}{(2\pi)^4}
 b(q^2)s(q^2)\Sigma_\pi(q^2)\frac{q^2(q^2\delta_{\mu\nu}-q_\mu q_\nu)}
{1+\Xi(q^2)}\omega^\mu(q)\rho^{0\nu}(-q).
\end{eqnarray}
Thus chain approximation of fig. 4 yield effective lagrangian as follow
\begin{eqnarray}\label{4.21}
{\cal L}_{\omega\rho}^*=\frac{N_c}{6\pi^2}\frac{m_u-m_d}{m}
 \int\frac{d^4q}{(2\pi)^4}e^{iq\cdot x}\{g^{-2}h_0(q^2)-
 \frac{q^2}{(1+\Xi(q^2))}b(q^2)s(q^2)\Sigma_\pi(q^2)\}
 (q^2\delta_{\mu\nu}-q_\mu q_\nu)\omega^\mu(q^2)\rho^{0\nu}(x).
\end{eqnarray}

\section{$\omega-\rho^0$ Mixing and $\omega\rightarrow\pi^+\pi^-$ Decay}

Due to VMD, the eleectromagnetic interaction contributes to
$\omega\rightarrow\pi^+\pi^-$ decay through 
$\omega\rightarrow\gamma\rightarrow\rho^0\rightarrow\pi\pi$ and 
$\omega\rightarrow\gamma\rightarrow\pi\pi$. In {\bf I} we have evaluated
$\rho\pi\pi$ vertex, $\rho^0-\gamma$ mixing vertex and ``direct''
$\gamma\pi\pi$ vertex up to one-loop level. The ``direct'' $\gamma\pi\pi$
vertex reads
\begin{eqnarray}\label{5.1}
{\cal L}_{\gamma\pi\pi}^c=\int\frac{d^4q}{(2\pi)^4}e^{iq\cdot x}
\bar{F}_\pi(q^2)A_\mu(q)[\pi^+(x)\pa^\mu\pi^-(x)-\pa^\mu\pi^+(x)\pi^-(x)], 
\end{eqnarray}
where $A_\mu$ is photon field, $\bar{F}_\pi(q^2)$ is nonresonant
background part of pion form factor. Explicitly, $\bar{F}_\pi(q^2)$ reads
\begin{eqnarray}\label{5.2}
\bar{F}_\pi(q^2)=1+\frac{q^2b_\gamma(q^2)}{1+\Sigma(q^2)},
\end{eqnarray}
where 
\begin{eqnarray}\label{5.3}
b_\gamma(q^2)&=&\frac{b(q^2)}{2(1+3\zeta)}-D(q^2)
  -\frac{C(q^2)\Sigma_0(q^2)}{1+11\zeta/3}, \nonumber \\
C(q^2)&=&\frac{1}{2f_\pi^2}[A(q^2)+2g_A^2B(q^2)],\nonumber \\
\Sigma_0(q^2)&=&\frac{2}{f_\pi^2}[2\Sigma_\pi(q^2)-\Sigma_K(q^2)],\\
\Sigma(q^2)&=&[1+\frac{q^2C(q^2)}{1+11\zeta/3}]\Sigma_0(q^2), 
  \nonumber \\
D(p^2)&=&\frac{1}{16\pi^2f_\pi^2}\{\lambda+\int_0^1dx\cdot
  x(1-x)\ln{[(1-\frac{x(1-x)p^2}{m_{_K}^2})
 (\frac{x(1-x)p^2}{m_{_K}^2})^2]}\nonumber \\&&\hspace{1in}
 -\frac{2}{3}i\pi\theta(p^2-4m_\pi^2)\}.
     \nonumber
\end{eqnarray}

The complete $\rho\pi\pi$ vertex reads
\begin{equation}\label{5.4}
{\cal L}_{\rho\pi\pi}^c=\int
  \frac{d^4q}{(2\pi)^4}e^{iq\cdot x}g_{\rho\pi\pi}(q^2)
  (q^2\delta_{\mu\nu}-q_\mu q_\nu)\rho^{0\mu}(q)
  [\pi^+(x)\pa^\mu\pi^-(x)-\pa^\mu\pi^+(x)\pi^-(x)],  
\end{equation}
with
\begin{equation}\label{5.5}
g_{\rho\pi\pi}(q^2)=\frac{b(q^2)}{(1+2\zeta)(1+\Sigma(q^2))}.
\end{equation}

Moreover, the complete $\rho-\gamma$ mixing vertex reads
\begin{equation}\label{5.6}
{\cal L}_{\rho\gamma}^c=-\frac{1}{2}e\int\frac{d^4q}{(2\pi)^4}
  e^{iq\cdot x}b_{\rho\gamma}(q^2)(q^2\delta_{\mu\nu}-q_\mu q_\nu)
  \rho^{0\mu}(q)A^\nu(x),
\end{equation}
where
\begin{equation}\label{5.7}
b_{\rho\gamma}(q^2)=\frac{A(q^2)}{g(1+\zeta)}-f_\pi^2b(q^2)
  \frac{\Sigma_0(q^2)}{1+2\zeta}[1
  +\frac{q^2b_\gamma(q^2)}{1+\Sigma(q^2)}].
\end{equation}
Thus due to VMD, the complete $\omega-\gamma$ mixing vertex can be
obtained directly
\begin{equation}\label{5.8}
{\cal L}_{\omega\gamma}^*=-\frac{1}{6}e\int\frac{d^4q}{(2\pi)^4}
  e^{iq\cdot x}b_{\rho\gamma}(q^2)(q^2\delta_{\mu\nu}-q_\mu q_\nu)
  \rho^{0\mu}(q)A^\nu(x).
\end{equation}
Eqs.(~\ref{5.6}) and (~\ref{5.8}) will lead to $\omega-\rho^0$ mixing at
the order of $\alpha_{\rm e.m.}$ through the transition process
$\omega\rightarrow\gamma\rightarrow\rho^0$, which is 
\begin{equation}\label{5.9}
{\cal L}_{\omega\rho}^{\rm e.m.}=\frac{1}{12}e^2\int
  \frac{d^4q}{(2\pi)^4}e^{iq\cdot x}b_{\rho\gamma}^2(q^2)
  (q^2\delta_{\mu\nu}-q_\mu q_\nu)\omega^\mu(q)\rho^{0\nu}(x).
\end{equation}
In addition, eqs.(~\ref{5.1}) and (~\ref{5.8}) also lead to ``direct''
$\omega\pi\pi$ couping at the order of $\alpha_{\rm e.m.}$ through the
transition process $\omega\rightarrow\gamma\rightarrow\pi\pi$, which is
\begin{eqnarray}\label{5.10}
{\cal L}_{\omega\pi\pi}^{\rm e.m.}&=&-\frac{i}{6}e^2\int
  \frac{d^4q}{(2\pi)^4}e^{iq\cdot x}\bar{F}_\pi(q^2)b_{\rho\gamma}(q^2)
  (q^2\delta_{\mu\nu}-q_\mu q_\nu) \nonumber \\&&\hspace{0.5in}\times
  \omega^\mu(q)[\pi^+(x)\pa^\nu\pi^-(x)-\pa^\nu\pi^+(x)\pi^-(x)].
\end{eqnarray}

Eq.(~\ref{5.9}) together with eqs.(~\ref{4.5}), (\ref{4.11}) and
(\ref{4.21}) give the complete $\omega-\rho^0$ mixing vertex as follow 
\begin{equation}\label{5.11}
{\cal L}_{\omega\rho}^{c}=\int\frac{d^4q}{(2\pi)^4}e^{iq\cdot x}
 \Theta_{\omega\rho}(q^2)(q^2\delta_{\mu\nu}-q_\mu q_\nu)
  \omega^\mu(q)\rho^{0\nu}(x),
\end{equation}  
where vector meson fields have been normalized to physical fields, and
\begin{eqnarray}\label{5.12}
\Theta_{\omega\rho}(q^2)=\frac{N_c}{6\pi^2}\frac{m_u-m_d}{m}
 \{g^{-2}h_0(q^2)(1-\frac{4}{3}\zeta)+q^2b(q^2)s(q^2)
   [\Sigma_K(q^2)-\frac{\Sigma_\pi(q^2)}{1+\Xi(q^2)}]\}
  +\frac{\alpha\pi}{3}b_{\rho\gamma}^2(q^2).
\end{eqnarray}

The complete ``direct'' $\omega\pi\pi$ vertex can be obtained via
summing eqs.(\ref{4.7}), (\ref{4.13}), (\ref{4.19}) and (\ref{5.10}),
\begin{eqnarray}\label{5.13}
{\cal L}_{\omega\pi\pi}^c&=&
  -i\int\frac{d^4q}{(2\pi)^4}e^{iq\cdot x}g_{\omega\pi\pi}(q^2)
  (q^2\delta_{\mu\nu}-q_\mu q_\nu) 
  \omega^\mu(q)[\pi^+(x)\pa^\nu\pi^-(x)-\pa^\nu\pi^+(x)\pi^-(x)].
\end{eqnarray}
where all meson fields have been normalized to physical fields, and
$g_{\omega\pi\pi}(q^2)$ is defined as follow
\begin{eqnarray}\label{5.14}
g_{\omega\pi\pi}(q^2)&=&\frac{N_c}{12\pi^2}\frac{m_u-m_d}{m}
  \{s(q^2)\left(\frac{1}{1+\Xi(q^2)}-\frac{10}{3}\zeta\right)
       \nonumber \\&&   
   -6f_\pi^{-2}\Sigma_K(q^2)[8m^2f_\pi^{-2}b(q^2)-\frac{s(q^2)}{3}
   (1+\frac{q^2N_c}{4\pi^2f_\pi^2})]\}
   +\frac{2\alpha\pi}{3}\bar{F}_\pi(q^2)b_{\rho\gamma}(q^2).
\end{eqnarray}

Thus G-parity forbidden $\omega\rightarrow\pi^+\pi^-$ includes a
nonresonant background contribution, eq.(~\ref{5.13}), and $\rho$
resonance exchange contribution(eqs.(~\ref{5.4}) and (\ref{5.11})). The
decay width on $\omega$ mass-shell is
\begin{eqnarray}\label{5.15}
\Gamma(\omega\rightarrow\pi^+\pi^-)=\frac{m_\omega^5}{48\pi}
|\frac{m_\omega^2\Theta_{\omega\rho}(m_\omega^2)
 g_{\rho\pi\pi}(m_\omega^2)}{m_\omega^2-m_\rho^2+im_\rho\Gamma_\rho}
  -g_{\omega\pi\pi}(m_\omega^2)|^2
 (1-\frac{4m_\pi^2}{m_\omega^2})^{3/2}.
\end{eqnarray}
Using the experimental data $B(\omega\rightarrow\pi^+\pi^-)=(2.21\pm
0.30)\%$\cite{PDG98} together with eq.(~\ref{5.15}), we obtain
\begin{equation}\label{5.16}
 m_u-m_d=-(3.9\pm 0.22){\rm MeV}
\end{equation}
at energy scale $\mu\sim m_\omega$. Here the error bar is from the
uncertainty in branch ratio of the process $\omega\rightarrow\pi^+\pi^-$.
In the standard way, the $\omega-\rho^0$ mixing amplitude is
\begin{eqnarray}\label{5.17}
\int\frac{d^4q}{(2\pi)^4}\Pi_{\omega\rho}(q^2)=
<\omega|\int d^4x{\cal L}_{\omega\rho}(x)|\rho>\; 
\Longrightarrow\Pi_{\omega\rho}(q^2)&=&q^2\Theta_{\omega\rho}(q^2).
\end{eqnarray}
The off-shell $\omega-\rho^0$ mixing amplitude is obviously momentum
dependent, and vanished at $q^2=0$. This is consistent with the arguement
by O'Connell {\sl et. al.} in ref.\cite{OC94} that this mixing amplitude
must vanish at the transition from time-like to space-like four momentum
within a broad class of models. In addition, the value of isospin broken
parameter(~\ref{5.16}) leads on $\omega$ mass-shell $\omega-\rho^0$ mixing
amplitude as follow
\begin{equation}\label{5.18}
{\rm Re}\Pi_{\omega\rho}(m_\omega^2)=-(3956\pm 280){\rm MeV}^2,
 \hspace{0.5in}
{\rm Im}\Pi_{\omega\rho}(m_\omega^2)=-(1697\pm 130){\rm MeV}^2.
\end{equation}
In ref.\cite{OC98}, the on-shell mixing amplitude has extracted from the
$e^+e^-\rightarrow\pi^+\pi^-$ experimental data in a model-dependent way.
In eq.(~\ref{5.18}), the real part of on-shell mixing amplitude 
agree with result of ref.\cite{OC98}. The imaginary part, however, is much
larger than one in ref.\cite{OC98} which is around $-300$MeV$^2$. It must
be pointed out that, in ref.\cite{OC98} the author's analysis bases on a
model without ``direct'' $\omega\pi\pi$ coupling. Therefore, it is
insignificant to compare the value of on-shell mixing amplitude of this
the present paper with one of ref.\cite{OC98}. Fortunately, the ratio
between $\omega\rightarrow\pi\pi$ decay amplitude and
$\rho\rightarrow\pi\pi$ decay amplitude should be model-indenpendent. This
value can test whether a model is right or not. The on-shell mixing
amplitude in ref.\cite{OC98} yields
\begin{equation}\label{5.19}
R_{\omega\rho}^{\rm exp}=\frac{<\pi^+\pi^-|\omega>}{<\pi^+\pi^-|\rho>}
       =-(0.0060\pm 0.0009)+(0.0322\pm 0.0050)i.
\end{equation}
The present paper predicts
\begin{equation}\label{5.20}
R_{\omega\rho}=\frac{m_\omega^2\Theta_{\omega\rho}(m_\omega^2)}  
 {m_\omega^2-m_\rho^2+im_\rho\Gamma_\rho}
  -\frac{g_{\omega\pi\pi}(m_\omega^2)}{g_{\rho\pi\pi}(m_\omega^2)}
  =-(0.0084\pm 0.0007)+(0.0331\pm 0.0021)i.
\end{equation}
We can see that this theoretical prediction agree with experimental
excellently.

Moreover, the follows have also been revealed in our studies of this
paper:

i) If we take $g_{\omega\pi\pi}(q^2)=0$ in eq.(~\ref{5.15}), we have
$B(\omega\rightarrow\pi^+\pi^-)=(2.56\pm 0.34)\%$. So that the
contribution from interference between ``direct'' $\omega\pi\pi$
coupling and $\omega-\rho^0$ mixing is about $15\%$. The dominant
contribution are from $\rho$-resonance exchange. This conclusion indicates
all pervious studies which without ``direct'' $\omega\pi\pi$ coupling are 
good approximation even though this neglect is an ad hoc assumption. 
However, in mechanism of $\omega\rightarrow\pi^+\pi^-$
with ``direct'' $omega\pi\pi$ coupling, larger imagnary part of
on-shell $\omega-\rho^0$ mixing amplitude is allowed, but it is not
allowed in the mechnism without the direct couping.

ii) If we do not consider the contributions from one-loop diagrams of
pseudoscalar mesons, i.e. setting $\Sigma_K(q^2)=\Sigma_\pi(q^2)=0$, we
obtain $B(\omega\rightarrow\pi^+\pi^-)=(2.86\pm 0.47)\%$. Thus the
contribution from one-loop of pseudoscalar mesons is about $30\%$ and can
not be omitted. This conclusion is consistent with {\bf I}. In addition,
in this case, the on-shell $\omega-\rho^0$ mixing amplitude is about
$-4700$MeV$^2$. So that we can see that the larger imagnary part of
on-shell $\omega-\rho^0$ mixing amplitude is yielded by pseudoscalar meson
loops. In {\bf I}, we have shown that this larger imagnary part is
required by the unitarity of this effective field theory.

\section{Summary}

In this paper, we study G-parity forbidden $\omega\rightarrow\pi^+\pi^-$
decay up to one-loop level of mesons. This process is yielded by isospin
symmetry breaking due to $m_u\neq m_d$ and electromagnetic interaction of
mesons. The decay amplitude contains two parts of contributions which are
from ``direct'' $\omega\pi\pi$ couping and $\omega-\rho^0$ mixing
respectively. In the previous studies, the ``direct'' $\omega\pi\pi$
couping is neglected. We show that the ``direct'' $\omega\pi\pi$ couping
and its interference with $\omega-\rho^0$ mixing contribute to
on-shell decay amplitude about 15$\%$ only. It also interprets why the
previous studies are good approximations even without ``direct''
$\omega\pi\pi$ couping. We suggest that the decay amplitude ratio
$R_{\omega\rho}$ should be model-independent, and our prediction agree
with experimental data excellently.

The formula of $\omega-\rho^0$ mixing amplitude is also obtained. Since
our calculation is beyond the chiral expansion(including all orders
contribution of the chiral expansion) and one-loop contribution of
pseudoscalar mesons is considered, the momentum-dependence of the
off-shell mixing amplitude is very complicated. However, the mixing
amplitude also vanishes at $q^2=0$. For case of on $\omega$ mass-shell,
the mixing amplitude emerges larger imagnary which is from one-loop
contribution of pion and is required by unitarity of this effective field
theory.

In our calculation, all vertices are expanded to the leading order light
current quark masses. At this order, the decay amplitude yielded by
isospin broken is proportional to $m_d-m_u$. The theorectical prediction
of isospin breaking parameter is $m_d-m_u=(3.9\pm 0.22)$MeV at energy
scale $\mu\sim m_\omega$. This value is important for determining light
quark masses at vector meson energy scale.
\begin{eqnarray*}
\\ 
\end{eqnarray*}

\begin{center}
\large{\bf Appendix}
\end{center}

Here we provide some identities which are used in calculation on
four-point effective action. In sect. 3.2 we have used $q$, $k$ and $-k-q$
to denote four-momentum square of external source $\bar{V}_\mu$,
$\Delta_\nu$ and $\Delta_\sigma$. For the purpose of this paper,
$\Delta_\nu(k)$ reduces to $k_\nu\pi(k)$. So that due to soft pion theorem
we have $k^2\rightarrow 0$, $(q+k)^2\rightarrow 0$ and
$k_\mu\Delta^\mu(k)\rightarrow 0$. Moreover, due to space-like condition
of vector meson fields, $q^\mu V_\mu(q)=0$, we have 
\begin{eqnarray}\label{A.1}
&&(\delta_{\mu\sigma}q_\nu+\delta_{\mu\nu}q_\sigma)
 <\{V^\mu(q),S\}\Delta^\nu(k)\Delta^\sigma(-k-q)>\;=\;0, \nonumber \\
&&(\delta_{\mu\sigma}q_\nu+\delta_{\mu\nu}q_\sigma)
 <V^\mu(q)\Delta^\nu(k)S\Delta^\sigma(-k-q)>\;=\;0.  \nonumber \\
&&q_\nu q_\sigma k_\mu<\{V^\mu(q),S\}
 \Delta^\nu(k)\Delta^\sigma(-k-q)>\;\rightarrow\;
 -\frac{q^2}{2}q_\sigma\delta_{\mu\nu}<\{V^\mu(q),S\}
 \Delta^\nu(k)\Delta^\sigma(-k-q)>, \nonumber \\
&&q_\nu q_\sigma k_\mu<V^\mu(q)\Delta^\nu(k)S\Delta^\sigma(-k-q)>
 \;\rightarrow\;-\frac{q^2}{2}q_\sigma\delta_{\mu\nu}<V^\mu(q)
 \Delta^\nu(k)S\Delta^\sigma(-k-q)>, \\
&&k_\mu\delta_{\nu\sigma}<\{V^\mu(q),S\}\Delta^\nu(k)
 \Delta^\sigma(-k-q)>\;\rightarrow\;-q_\sigma\delta_{\mu\nu}
 <\{V^\mu(q),S\}\Delta^\nu(k)\Delta^\sigma(-k-q)>,  \nonumber \\
&&k_\mu\delta_{\nu\sigma}
 <V^\mu(q)\Delta^\nu(k)S\Delta^\sigma(-k-q)>
 \;\rightarrow\;-q_\sigma\delta_{\mu\nu}
 <V^\mu(q)\Delta^\nu(k)S\Delta^\sigma(-k-q)>, \nonumber 
\end{eqnarray}
The following integral identities are also used in our calculation
\begin{eqnarray}\label{A.2}
&&\int_0^1dx\cdot x^2\int_0^1dy(1-y)\frac{xy}{[l^2-m^2+xy(1-x)q^2]^4}
=\int_0^1dx\cdot x^2\int_0^1dy(1-y)\frac{(1-x)}
  {[l^2-m^2+xy(1-x)q^2]^4} \nonumber \\
&=&\int_0^1dx\cdot x^2\int_0^1dy(1-y)\frac{x(1-2y)}
 {[l^2-m^2+x^2y(1-x)q^2]^4} 
=\int_0^1dx\cdot x^2\int_0^1dy(1-y)\frac{(1-x)}
 {[l^2-m^2+x^2y(1-x)q^2]^4}.
\end{eqnarray}

\begin{center}
{\bf ACKNOWLEDGMENTS}
\end{center}
This work is partially supported by NSF of China through C. N.
Yang and the Grant LWTZ-1298 of Chinese Academy of Science.

\end{document}